\documentclass[fleqn,twoside]{article}
\usepackage{espcrc2}

\usepackage{graphicx}
\usepackage[figuresright]{rotating}

\newcommand{\AmS}{{\protect\the\textfont2
  A\kern-.1667em\lower.5ex\hbox{M}\kern-.125emS}}

\title{Neutrino Mass Models: circa 2008\thanks{Based on Plenary Talk presented at the Neutrino Oscillation Workshop (NOW2008), Conca Specchiulla, Italy, September 6-13, 2008.}}

\author{Mu-Chun Chen\address{Department of Physics \& Astronomy, University of California, Irvine, CA 92697-4575, USA}%
        ~and
        K.T. Mahanthappa\address{Department of Physics, University of Colorado, Boulder, CA 80309-0390, USA}%
       }
\begin{document}

\begin{abstract}
We review recent developments in theoretical models for neutrino masses and mixing. Emphases are given to models based on finite group family symmetries. In particular, we describe one recent model based on $SU(5)$, in which both the tri-bimaximal neutrino mixing and realistic CKM matrix are generated.  We also discuss two models based on a non-anomalous $U(1)_{F}$ family symmetry in which the gauge anomalies are cancelled due to the presence of the right-handed neutrinos. In one of these models, the seesaw scale can be as low as a TeV; in the other model, which is based on SUSY $SU(5)$, the $U(1)_{F}$ symmetry forbids Higgs-mediated proton decays.    
\vspace{1pc}
\end{abstract}

\maketitle

\section{INTRODUCTION}

The recent  advent of the neutrino oscillation data from Super-Kamiokande has provided a solid evidence that neutrinos have small but non-zero masses. The global fit to current data from neutrino oscillation experiments give the following best fit values and $2\sigma$ limits for the mixing parameters~\cite{Maltoni:2004ei},
\begin{eqnarray}
\sin^{2} \theta_{12} & = & 0.30 \; (0.25 - 0.34),  \nonumber \\
\sin^{2} \theta_{23} & = &  0.5 \; (0.38 - 0.64), \nonumber \\
\sin^{2} \theta_{13} & = & 0 \;  (< 0.028), \nonumber \\
\Delta m_{12}^{2} & = & 7.9 \; (7.3 - 8.5) \; \mbox{eV}^2, \nonumber\\
\Delta m_{23}^{2} & = & 2.2 \; (1.7 - 2.9) \; \mbox{eV}^2. \nonumber
\end{eqnarray}
In addition, recent analyses~\cite{Fogli:2008jx} from the Bari group have given hints on possible non-zero value for $\theta_{13}$, with 
\begin{displaymath}
\sin^{2}\theta_{13} = 0.016 \pm 0.010 \; , 
\end{displaymath}
at $1 \, \sigma$.   
Since then, the measurements of neutrino oscillation parameters have entered a precision era. In the Standard Model, due to the lack of  right-handed neutrinos and lepton number conservation, neutrinos are massless. To generate non-zero neutrino masses thus calls for physics beyond the Standard Model. There have been many theoretical ideas proposed with an attempt to accommodate the experimentally observed  small neutrino masses and the larger mixing angles among them. Most of the models are either based on grand unification combined with family symmetries or having family symmetries in the lepton sector only.  Recently, it was realized that small neutrino masses can also arise with new physics at the TeV scale, contrary to the common belief that the scale of the seesaw mechanism has to be high. 
In this talk, we review some of these ideas as well as the predictions of various existing models. For more extensive reviews, see, for example, Ref.~\cite{Chen:2003zv}. 

\begin{table*}[htb]
\caption{Charge assignments. Here the parameter $\omega = e^{i\pi/6}$.}  
\label{tbl:charge} 
\vspace{0.06in}
\begin{tabular}{|c|ccc|ccc|cccccc|cc|}\hline
& $T_{3}$ & $T_{a}$ & $\overline{F}$ & $H_{5}$ & $H_{\overline{5}}^{\prime}$ & $\Delta_{45}$ & $\phi$ & $\phi^{\prime}$ & $\psi$ & $\psi^{\prime}$ & $\zeta$ & $N$ & $\xi$ & $\eta$  \\ [0.3em] \hline\hline
SU(5) & 10 & 10 & $\overline{5}$ & 5 &  $\overline{5}$ & 45 & 1 & 1 & 1 & 1& 1 & 1 & 1 & 1\\ \hline
${ }^{(d)}T$ & 1 & $2$ & 3 & 1 & 1 & $1^{\prime}$ & 3 & 3 & $2^{\prime}$ & $2$ & $1^{\prime\prime}$ & $1^{\prime}$ & 3 & 1 \\ [0.2em] \hline
$Z_{12}$ & $\omega^{5}$ & $\omega^{2}$ & $\omega^{5}$ & $\omega^{2}$ & $\omega^{2}$ & $\omega^{5}$ & $\omega^{3}$ & $\omega^{2}$ & $\omega^{6}$ & $\omega^{9}$ & $\omega^{9}$ 
& $\omega^{3}$ & $\omega^{10}$ & $\omega^{10}$ \\ [0.2em] \hline
$Z_{12}^{\prime}$ & $\omega$ & $\omega^{4}$ & $\omega^{8}$ & $\omega^{10}$ & $\omega^{10}$ & $\omega^{3}$ & $\omega^{3}$ & $\omega^{6}$ & $\omega^{7}$ & $\omega^{8}$ & $\omega^{2}$ & $\omega^{11}$ & 1 & $1$ 
\\ \hline   
\end{tabular}
\vspace{0.1in}
\end{table*}

\section{FINITE GROUP FAMILY SYMMETRIES}

The experimental best fit values for the mixing parameters are very close to the values arising from the so-called ``tri-bimaximal'' mixing (TBM) matrix~\cite{Harrison:2002er},
\begin{equation}
U_{\mathrm{TBM}} = \left(\begin{array}{ccc}
\sqrt{2/3} & 1/\sqrt{3} & 0\\
-\sqrt{1/6} & 1/\sqrt{3} & -1/\sqrt{2}\\
-\sqrt{1/6} & 1/\sqrt{3} & 1/\sqrt{2}
\end{array}\right) \; , \label{eq:tri-bi}
\end{equation}
which predicts 
\begin{eqnarray}
\sin^{2}\theta_{\mathrm{atm, \, TBM}} & = & 1/2 \; ,  \nonumber\\
\sin^{2}\theta_{\odot, \mathrm{TBM}} & = & 1/3 \; , \nonumber\\
\sin\theta_{13, \mathrm{TBM}} & = & 0\; .
 \end{eqnarray}
Even though the predicted $\theta_{\odot, \mathrm{TBM}}$ is currently still  allowed by the experimental data at $2\sigma$, as it is very close to the upper bound at the $2\sigma$ limit, it  may be ruled out once more precise measurements are made in the  upcoming experiments.  

The tri-bimaximal neutrino mixing pattern can arise if the neutrino mass matrix has the following form,
\begin{equation}
M_{\nu} = \left(
\begin{array}{ccc}
a & b & b \\
b & c & d \\
b & d & c
\end{array}\right) \; .
\end{equation}
This matrix predicts $\sin^{2} 2 \theta_{23} = 1$ and $\theta_{13}=0$, while leaving the value for $\theta_{12}$ undetermined. This mass matrix can arise from an 
underlying $S_{3}$~\cite{Mohapatra:2006pu}, $D_{4}$~\cite{Grimus:2003kq}, or $\mu-\tau$ symmetry~\cite{Fukuyama:1997ky}. A prediction for $\tan^{2}\theta_{12} = 1/2$ arises if  the parameters are chosen such that $a+b = c+d$ is satisfied. 

It has been pointed out that the tri-bimaximal mixing matrix can arise from a family symmetry in the lepton sector based on $A_{4}$~\cite{Ma:2001dn}, which automatically gives rise to, $a+b = c+d$, leading to a prediction for the solar mixing angle, $\sin^{2}\theta_{12} = 1/3$. However, due to its lack of doublet representations, CKM matrix is an identity in most $A_{4}$ models. It is hence not easy to implement $A_{4}$ as a family symmetry for both quarks and leptons~\cite{Ma:2006sk}.

\subsection{A Realistic $SU(5)\times { }^{(d)}T$ Model}
\vspace{0.05in}

In \cite{Chen:2007af}, a grand unified model based on SU(5) combined with the double tetrahedral group~\cite{Frampton:1994rk}, ${}^{(d)}T$, was constructed, which successfully gives rise to near tri-bimaximal leptonic mixing as well as realistic CKM matrix elements for the quarks. The group ${}^{(d)}T$ is the double covering group of $A_{4}$. In addition to the $1, \; 1^{\prime}, \; 1^{\prime\prime}$ and $3$ representations that $A_{4}$ has, the group ${}^{(d)}T$ also has three in-equivalent doublet representations, $2, \; 2^{\prime}, \; 2^{\prime\prime}$. This enables the $(1+2)$ assignments, which has been shown to give realistic masses and mixing pattern in the quark sector~\cite{so10ref}.   
The charge assignments of various fields are summarized in Table~\ref{tbl:charge}.  
Due to the presence of the  
$Z_{12} \times Z_{12}^{\prime}$ symmetry, only nine operators are allowed in the model, and hence the model is very predictive, the total number of parameters being nine in the Yukawa sector for the charged fermions and the neutrinos. The Lagrangian of the model is given as follows,
\begin{eqnarray}
\mathcal{L}_{\mathrm{Yuk}} &  = &   \mathcal{L}_{\mathrm{TT}} + \mathcal{L}_{\mathrm{TF}} + \mathcal{L}_{\mathrm{FF}} \\
\mathcal{L}_{\mathrm{TT}} &= & y_{t} H_{5} T_{3}T_{3} + \frac{1}{\Lambda^{2}} y_{ts} H_{5} T_{3} T_{a} \psi \zeta \nonumber
\\  &&\hspace{-0.2in} + \frac{1}{\Lambda^{2}} y_{c} H_{5} T_{a} T_{a} \phi^{2} + \frac{1}{\Lambda^{3}} y_{u} H_{5} T_{a} T_{a} \phi^{\prime 3}   \nonumber \\  
\mathcal{L}_{\mathrm{TF}}  &=&  \frac{1}{ \Lambda^{2}}  y_{b} H_{\overline{5}}^{\prime} \overline{F} T_{3} \phi \zeta \nonumber \\
 &&\hspace{-0.26in} + \frac{1}{\Lambda^{3}} \biggl[ y_{s} \Delta_{45} \overline{F} T_{a} \phi \psi N  + 
 y_{d} H_{\overline{5}}^{\prime} \overline{F} T_{a} \phi^{2} \psi^{\prime}  \biggr] \nonumber \\
\mathcal{L}_{\mathrm{FF}} & =&  \hspace{-0.15in} \frac{1}{M_{x}\Lambda} \biggl[\lambda_{1} H_{5} H_{5} \overline{F} \, \overline{F} \xi +  \lambda_{2} H_{5} H_{5} \overline{F} \, \overline{F} \eta\biggr] \; , \nonumber
 \end{eqnarray}
 where $M_{x}$ is the cutoff scale at which the lepton number violation operator $HH\overline{F}\, \overline{F}$ is generated, while $\Lambda$ is the cutoff scale, above which the ${}^{(d)}T$ symmetry is exact.  (For the VEV's of various scalar fields, see Ref.~\cite{Chen:2007af}.) The parameters $y$'s and $\lambda$'s are the coupling constants. 
 
 The interactions in $\mathcal{L}_{FF}$ give the following neutrino mass matrix, 
\begin{displaymath}
M_{\nu} = 
\frac{\lambda v^{2}}{M_{x}} \left(\begin{array}{ccc}
2 \xi_{0} + u & -\xi_{0} & -\xi_{0} \\
-\xi_{0} & 2 \xi_{0} & u - \xi_{0} \\
-\xi_{0} & u-\xi_{0} & 2 \xi_{0}\end{array}\right) \; , 
\end{displaymath} 
and we have absorbed the Yukawa coupling constants by rescaling the VEV's.  
This mass matrix $M_{\nu}$ is form diagonalizable, {\it i.e.} the orthogonal matrix that diagonalizes it does not depend on the eigenvalues. Its diagonal form is,
\begin{equation}
V_{\nu}^{\mathrm{T}} M_{\nu} V_{\nu} = \mathrm{diag}(u+ 3\xi_{0}, \, u, \, -u + 3\xi_{0}) \frac{v^{2}_{u}}{M_{x}} \; ,
\end{equation}
where the matrix $V_{\nu}$ is the tri-bimaximal mixing matrix, $V_{\nu} = U_{\mathrm{TBM}}$.  
 
Due to the $Z_{12}$ symmetry,  the mass hierarchy arises dynamically without invoking an additional U(1) symmetry. The $Z_{12}$ symmetry also forbids Higgsino-mediated proton decays in SUSY version of the model. Due to the ${ }^{(d)}T$ transformation property of the matter fields, the $b$-quark mass can be generated only when the ${ }^{(d)}T$ symmetry is broken, which naturally explains  the hierarchy between $m_{b}$ and $m_{t}$. 
The $Z_{12} \times Z_{12}^{\prime}$ symmetry, to a very high order, also forbids operators that lead to nucleon decays. In principle, a symmetry smaller than  $Z_{12} \times Z_{12}^{\prime}$ would suffice in getting realistic masses and mixing pattern; however, more operators will be allowed and the model would not be as predictive.  
The Georgi-Jarlskog relations for three generations are obtained. This inevitably requires non-vanishing mixing in the charged lepton sector, leading to corrections to the tri-bimaximal mixing pattern. The model predicts non-vanishing $\theta_{13}$, which is related to the Cabibbo angle as, 
\begin{equation}
\theta_{13}\sim \theta_{c}/3\sqrt{2}\; .
\end{equation}  
Numerically, this is close to $\sin\theta_{13} \sim 0.05$ which may be probed by the Daya Bay reactor experiment. In addition, it gives rise to a sum rule, 
\begin{equation}
\tan^{2}\theta_{\odot} \simeq \tan^{2} \theta_{\odot, \mathrm{TBM}} - \frac{1}{2} \theta_{c} \cos\delta \; ,
\end{equation} 
which is a consequence of the Georgi-Jarlskog relations in the quark sector (with $\delta$ being the Dirac CP phase in the lepton sector).\footnote{Such relation for the solar mixing angle is quite generic and was also found in a model based on the Pati-Salam group~\cite{King:2005bj}.} This deviation could account for the difference between the experimental best fit value for the solar mixing angle and the value predicted by the tri-bimaximal mixing matrix. 

Since the three absolute neutrino mass eigenvalues are determined by only two parameters, {\it i.e.} the VEVs $u_{0}$ and $\xi_{0}$, there is a sum rule that relates the three light masses, 
\begin{equation}
m_{1} - m_{3} = 2m_{2} \; .
\end{equation}
More generally, the three absolute masses can be complex,
\begin{eqnarray}
m_{1} & = & u_{0} + 3 \xi_{0} e^{i\theta} \; , \\
m_{2} & = & u_{0} \; ,  \\
m_{3} & = & -u_{0} + 3\xi_{0} e^{i\theta} \; 
\end{eqnarray} 
with $u_{0}$ and $\xi_{0}$ being real. 
It then follows the sum rule,
\begin{equation}
\Delta m_{\odot}^{2} = -9 \xi_{0}^{2} + \frac{1}{2} \Delta m_{atm}^{2} \; .
\end{equation}
Given that $\Delta m_{\odot}^{2} > 0$ is required in order to have matter effects in solar neutrino oscillation, it immediately follows from the above sum rule that the normal hierarchy pattern with $\Delta m_{atm}^{2} > 0$ is predicted.\footnote{See, also Ref.~\cite{Ma:2005sha} for a more general discussion on mass ordering.} 

\subsection{Comments on Leptogenesis}
\vspace{0.05in}

Since the exact tri-bimaximal neutrino mixing pattern predicts $\theta_{13}=0$, one question that arises is whether  leptogenesis is possible. (For a review on leptogenesis, see {\it e.g.} Ref.~\cite{Chen:2007fv}.) It was pointed out~\cite{Jenkins:2008rb} that for models that predict exact TBM pattern for neutrino mixing from an underlying family symmetry without any tuning, leptogenesis vanishes. This is true even when the flavor effects are included, due to the fact that there is no right-handed mixing in models with exact TBM neutrino mixing~\cite{Jenkins:2008rb}. Sufficient amount of leptogenesis can be generated once corrections due to higher dimensional operators to the exact tri-bimaximal mixing pattern are included.

In an $S_{3}$ model~\cite{Mohapatra:2006se} with Type-II seesaw mechanism in which the tri-bimaximal neutrino mixing is accommodated, non-vanishing  leptogenesis can be generated and its value is related to one of the Majorana phase.\footnote{In minimal left-right model with spontaneous CP violation, all leptonic CP violations, including those in leptongenesis and neutrino oscillation, are due to a single phase~\cite{Chen:2004ww}.}

\begin{table*}[tbh!]
\caption{$U(1)_F$ charges that satisfy the anomaly cancellation conditions while giving rise to realistic fermion masses and mixing angles in a SUSY $SU(5)$ model~\cite{Chen:2008tc}.}\label{table:2}
\vspace{0.06in}
\begin{tabular}{|c|c| c| c| c| c| c| c|c|c|c|c|}
\hline 
Field &${\bf \overline{5}}_1$&${\bf \overline{5}}_2$&${\bf
\overline{5}}_3$&${\bf 10}_1$&${\bf 10}_2$&${\bf
10}_3$&$N_1$&$N_2$&$N_3$&${\bf 5}_{H1}$&${\bf 5}_{H2}$\\
\hline
$U(1)_F$ 
&$1/2$&$-1/2$&$-1/2$&$25/18$&
$7/18$&$-29/18$&$59/18$&$5/18$
&$-49/18$&$29/9$&$-19/9$\\
Charge & & & & & & & & & & &\\
\hline
\end{tabular}
\end{table*}

\section{TEV SCALE SEESAW MECHANISM}

In the conventional wisdom, the smallness of the neutrino masses is tied to the high scale of the new physics that generates neutrino masses. As the new physics scale is high, it is very hard, if not impossible, to probe such new physics at current collider experiments. In \cite{Chen:2006hn}, an alternative was proposed in which the small neutrino masses are generated with TeV scale physics. This allows the possibility of testing the new physics that gives rise to neutrino masses at the Tevatron and the LHC. This is achieved by augmenting the Standard Model with a non-anomalous $U(1)_{\nu}$ symmetry and $N$ right-handed neutrinos. Due to the presence of  the $U(1)_{\nu}$ symmetry, neutrino masses can only be generated by operators with very high dimensionality, which in turn allows a low cut-off scale.

The new anomaly cancellation conditions
are highly non-trivial, especially because all fermion
charges are expected to be commensurate. Nevertheless, assuming that all quark Yukawa couplings
and all diagonal charged-lepton Yukawa couplings to the standard model Higgs doublet $H$ are gauge invariant,
 it is found that the most general solution to the anomaly cancellation conditions when
$N=1$ or 3. Only in the  $N=3$ case, scenarios consistent with
light neutrino masses and
$\Lambda$ at the TeV scale were found. 
For $N=3$, the charges of all quarks and leptons
(including right-handed neutrinos) are determined in terms of four rational parameters,
assumig one of the fermion charges is fixed by an appropriate normalization of the gauge
coupling. 

There exist regions in the parameter space that fit the neutrino oscillation data.
Depending on the choice of parameters, the neutrinos can be  either
Dirac or Majorana fermions. In scenarios with Majorana neutrinos, the existence of
``quasi-sterile'' neutrinos that mix slightly with the active neutrinos and couple to the new
$Z^{\prime}$ gauge boson is predicted. These quasi-sterile neutrinos may have interesting
phenomenological consequences for cosmology and oscillation physics. In the case of Dirac neutrinos, 
potentially observable consequences of the new degrees of freedom are also predicted.

Because the $U(1)_{\nu}$ symmetry is spontaneously broken at around the weak scale, 
the $Z^{\prime}$ gauge boson and the particles from the $U(1)_{\nu}$ breaking sector
will manifest themselves in a variety of interesting ways.  
$Z^{\prime}$ exchange can mediate neutral-fermion flavor
violating processes, which may be observable in next-generation 
neutrino oscillation experiments. The new heavy states can be discovered
in current and upcoming collider experiments, enabling the possibility of probing the neutrino sector at the collider experiments. 

\section{NON-ANOMALOUS $U(1)_{F}$ SYMMETRY in SUSY GUT}

It was claimed~\cite{Ibanez:1994ig} that in SUSY GUT models, in order to generate realistic fermion masses and mixing angles from a $U(1)_{F}$ family symmetry, the gauge anomalies can only be cancelled by the Green-Schwarz mechanism, and thus the $U(1)_{F}$ must be anomalous. However, this is not a no-go theorem, and in the model described below, solutions that satisfy all the gauge anomaly cancellation conditions and at the same time accommodate realistic fermion mass patterns have been found to exist.

\subsection{Non-anomalous $U(1)_{F}$ in a SUSY $SU(5)$ Model}
\vspace{0.05in}

In Ref.~\cite{Chen:2008tc}, a non-anomalous $U(1)_{F}$ was proposed as a family symmetry, in a SUSY $SU(5)$ model. There are three anomaly cancellation conditions that must be satisfied: the $[SU(5)]^{2}-U(1)$ mixed anomaly, the $U(1)$-gravitation anomaly, and the $[U(1)]^{3}$ anomaly. The set of charges that satisfy these conditions are given in Table~\ref{table:2}. These charges give rise to the following mass matrices,
\begin{eqnarray}
Y_{u} & \sim & \left(\begin{array}{ccc}
\lambda^{6} & \lambda^{5} & \lambda^{3} \\
\lambda^{5} & \lambda^{4} & \lambda^{2} \\
\lambda^{3} & \lambda^{2} & 1
\end{array}\right)\; , \\
Y_{d} & = & Y_{e}^{T} \sim \left(\begin{array}{ccc}
\lambda^{4} & \lambda^{3} & \lambda^{3} \\
\lambda^{3} & \lambda^{2} & \lambda^{2} \\
\lambda & 1 & 1
\end{array}\right) \; .
\end{eqnarray}
where $\lambda \equiv \frac{\left< \phi \right>}{\Lambda}$ with $\phi$ being the scalar field that breaks the $U(1)_{F}$ family symmetry.  
For the neutrinos,
\begin{eqnarray}
Y_{\nu} & \sim & \left(\begin{array}{ccc}
\lambda^{7} & \lambda^{4} & \lambda \\
\lambda^{6} & \lambda^{3} & 1 \\
\lambda^{6} & \lambda^{3} & 1
\end{array}\right)\; , \\
M_{RR} &  \sim & 
\left(\begin{array}{ccc}
\lambda^{6} & \lambda^{3} & 1 \\
\lambda^{3}  & 1 & \lambda^{3} \\
1 & \lambda^{3} & \lambda^{6}
\end{array}\right) \left< \chi \right>  \; ,
\end{eqnarray}
leading to the following effective neutrino mass matrix,
\begin{equation}
m_{\nu} \sim \lambda^{6} \left(\begin{array}{ccc}
\lambda^{2} & \lambda & \lambda \\
\lambda & 1 & 1 \\
\lambda & 1 & 1
\end{array}\right) \frac{v^{2}}{\left< \chi \right>}\; .
\end{equation}

It is interesting to note that, the non-anomalous $U(1)_{F}$ in this model also forbids operators that lead to proton decays mediated by the color-triplet Higgsinos.

\section{MODEL PREDICTIONS FOR THE OSCILLATION PARAMETERS}

\begin{figure}[tbh]
 \caption{Predictions for $\theta_{13}$ from various models~\cite{Albright:2006cw}.}\label{fig:all}
 \vspace{0.1in}
 \includegraphics[height=.3\textheight]{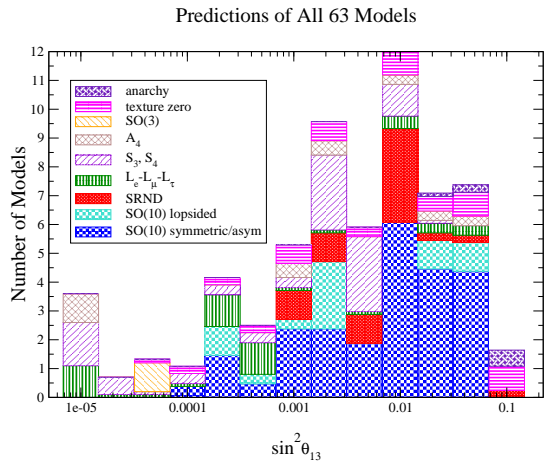}
\end{figure}

In \cite{Albright:2006cw}, a comparison of the predictions of some sixty-three models was presented.  
These include models based on $SO(10)$, 
models that utilize single RH neutrino dominance mechanism, 
and models based on family symmetries such as $L_{e} - L_{\mu} - L_{\tau}$ symmetry, 
$S_{3}$ symmetry, 
$A_{4}$ symmetry, 
and $SO(3)$ symmetry, 
as well as models based on texture zero assumptions. 
The predictions of these models for $\sin^{2}\theta_{13}$ are summarized in Fig.~\ref{fig:all}.    An observation one can draw immediately is that predictions of  $SO(10)$ models are larger than $10^{-4}$, and the median value is roughly $\sim 10^{-2}$.  Furthermore, $\sin^{2}\theta_{13} < 10^{-4}$ can only arise in models based on leptonic symmetries. However, these models are not as predictive as the GUT models, due to the uncertainty in the charged lepton mixing matrix. In this case, to measure $\theta_{13}$ will require a neutrino superbeam or a neutrino factory. In addition to the value of $\theta_{13}$, predictions for various LFV charged lepton processes can also be a powerful way to distinguish different models~\cite{Albright:2008ke}.

\vspace{-0.5in}
\vspace{-0.2in}

\section{CONCLUSIONS}

We review a few recent models for neutrino masses and mixing. We present a successful recent attempt based on a SU(5) grand unified model combined with ${}^{(d)}T$ symmetry, in which both the tri-bimaximal neutrino mixing and realistic CKM mixing matrix are generated.    A model based on non-anomalous $U(1)_{F}$ in which small neutrino masses are generated with new physics at the TeV scale has also been shown.  We also describe a SUSUY SU(5) model with non-anoumalous $U(1)_{F}$ for fermion masses. A study of existing models indicates that the range of predictions  for $\theta_{13}$ is very broad, although there are some characteristic model predictions with which more precise experimental measurements may tell different models apart.

\section{ACKNOWLEDGMENTS}
It is a great pleasure to thank the organizers, in particular Professors Gian Luigi Fogli and Eligio Lisi, for hosting such a stimulating workshop at the beautiful Conca seaside, for the hospitality they had extended throughout the meeting, and for their kind invitation to speak at the workshop. Thanks are also due to Carl Albright, Andre deGouvea, Bogdan A. Dobrescu, D. R. Tim Jones, Arvind Rajaraman and Hai-Bo Yu for their collaborations on some of the works presented here.  The work of M-CC was supported, in part, by the National Science Foundation under grant No. PHY-0709742. The work of KTM was supported, in part, by the Department of Energy under grant No.  DE-FG02-04ER41290.

\end{document}